\documentclass{article}

\usepackage[utf8]{inputenc}
\usepackage{natbib}
\usepackage{url}
\usepackage{graphicx}	
\usepackage{amsmath}	
\usepackage{amssymb}	

\usepackage{aas_macros}

\title{The Formation of Habitable Planets in the Four-Planet System HD~141399}

\author{R. Dvorak$^{1}$, B. Loibnegger$^{1}$, L.Y. Zhou$^{2}$, and L. Zhou$^{2}$
\\
$^{1}$Department of Astrophysics, University of Vienna, \\ A-1180 Vienna, Austria\\
$^{2}$School of Astronomy \& Space Science, Nanjing University, \\          163 Xianlin Avenue, Nanjing 210046, China
}

\begin{document}

\maketitle

\begin{abstract}
The presented work investigates the possible formation of 
terrestrial planets in the habitable zone (HZ) of the exoplanetary system HD~141399. In this system the HZ is located approximately between the planets {\bf c} (a = 0.7~au) and {\bf d} (a = 2.1~au). Extensive numerical integrations of the equations of motion in the pure Newtonian framework of small bodies with different initial conditions in the HZ are performed. Our investigations included several steps starting with 500 massless bodies distributed between planets {\bf c} and {\bf d} in order to model the development of the disk of small bodies. It turns out that after some $10^6$ years a belt-like structure analogue to the main belt inside Jupiter in our Solar System appears. We then proceed with giving the small bodies masses ($\sim$ Moon-mass) and take into account the gravitational interaction between these planetesimal-like objects. The growing of the objects -- with certain percentage of water -- due to collisions is computed in order to look for the formation of terrestrial planets. We observe that planets form in regions connected to mean motion resonances (MMR). So far there is no observational evidence of terrestrial planets in the system of HD~141399 but from our results we can conclude that the formation of terrestrial planets -- even with an appropriate amount of water necessary for being habitable -- in the HZ would have been possible.
\end{abstract}

\section{Introduction}

The 'race' to find a second Earth led to the discovery of many such candidates with different techniques from space and from ground. Nevertheless, although we have knowledge of systems hosting terrestrial\footnote{using the definition of a terrestrial planet meaning a planet smaller than 5 Earth-masses with a rocky surface} planets without gas giants -- for example Trappist~1 \footnote{7
terrestrial planets moving partly in the HZ, see \url{http://exoplanet.eu/}} \citep[e.g.,][]{Gillon2016} -- our Solar System is an example of hosting 
big planets outside of the habitable zone (HZ). This zone was proposed by \cite{Kasting1993} and later revised by e.g., \cite{Kopparapu2013} and describes the zone where water can exist in liquid form
on a terrestrial planet orbiting the star. 

Since the first discovery of a big planet in 1995 \citep{Mayor1995} -- a so called hot Jupiter --
many gas giants have been found around stars of different spectral type. 

The discoveries of many hot Jupiters over the past 20 years and the implications of their existence for the process of planet formation make it difficult to believe that terrestrial planets may co-exist in the HZ of those systems \citep{Dawson2018}. 

Systems with only one gas giant are well represented in the long list of extrasolar planets\footnote{see \url{http://exoplanet.eu/}}, but in our work we concentrate on systems with more than one gas giant of which several ones are known (see Table~\ref{Table1}). Of special interest to us seems the recently discovered four-planet system HD~141399 \citep{Vogt2014}. 

\begin{table}
\begin{center}

\begin{tabular}{l|lccc}
System & $\mathrm{m_{star}}$ & nr of & nr with  & m $>$ 0.8 $\mathrm{m_{Jup}}$ \\
& ($\mathrm{m_{Sun}}$) & Planets & m $>$ 0.8 $\mathrm{m_{Jup}}$ &  a $>$ 2.0~au \\
\hline
HD 125612 & 1.1   & 3 & 2 & 1 \\ 
HD 204313 & 1.045 & 3 & 2 & 2  \\ 
47 UMa    & 1.03  & 3 & 2 & 1 \\ 
GJ 676 A  & 0.71  & 4 & 2 & 1 \\ 
mu Ara    & 1.08  & 4 & 2 & 1 \\ 
ups And   & 1.27  & 4 & 3 & 2 \\ 
HR 8799   & 1.56  & 4 & 4 & 4 \\ 
55 Cnc	  & 0.95  & 5 & 2 & 1 \\ 
HD 131499 & 1.07  & 4 & 2 & 1 \\

\hline
\end{tabular}
\caption{A few examples of confirmed planetary systems with at least one gas giant at an orbit with a $>$ 2.0~au (check \protect\url{https://www.exoplanet.eu/} for updates and more detailed information). The system of 47~UMa is topic of another paper within our group \citep{Cuntz2018}.} \label{Table1}

\end{center}
\end{table}

\cite{Vogt2014} carefully describe the method of determining the
orbital elements of the planets and check their stability which seems to be
guaranteed because of the almost circular orbits of the four planets.

The aim of our research is to find whether terrestrial planets could be have been formed in the
HZ around the late K-type star. 

The structure of this research paper is as
follows: In Section~\ref{sec:system} we describe the system of HD~141399 in detail. The method of investigation is presented in Section~\ref{sec:method}. The results of the integration of massless bodies
in the HZ between planet {\bf c} and {\bf d} are shown in Section~\ref{sec:belt}, where we present a thorough analysis of
the outcomes and show examples of places where terrestrial planets can form on stable orbits. The results were gained using extensive
numerical simulations with integration times of up to 10~Myr. In Sub - section~\ref{sec:largemass} we take into account that the observations were made using radial velocity measurements which yield only a minimum mass for the planets and concetrate on the formation of bodies in the system with assumed larger masses for the giant planets.

Section~\ref{sec:examples} shows the results of computations with massive bodies which formed terrestrial planets through collisions.
In Section~\ref{sec:growth} we describe the growth process from planetesimals to protoplanets in more detail.  
Finally in Sections~\ref{sec:results} and \ref{sec:conclusion} we show the main results and draw the conclusions with focus on possible terrestrial planets in the HZ.

In Appendix~\ref{sec:dynmap} we investigate the dynamics around the Lagrangian Point $L_4$ of planet {\bf d} in detail and describe the acting secular resonances in the system in Appendix~\ref{sec:resonance}. We note that the Trojan region of planet {\bf d} is just outside of the HZ. In the presented article we included the computations for possibly habitable planets formed in this region taking into account the eccentricity of planet {\bf d}. On its orbit it will sometimes be inside and sometimes just outside the defined HZ.

\section{The System of HD~141399}\label{sec:system}

It is important to have an overview of possibly stable terrestrial planets in known multiple planetary 
systems with gas giants \citep{Agnew2018}. Nevertheless, we think that each system should be treated separately because of the different 
architecture. The dynamical stability of additional -- possibly terrestrial -- planets in a planetary
systems with gas giants is only one side of the coin, the other side is to know
how these planets could have formed. The present knowledge is that in our Solar System the gas giants formed prior to the terrestrial planets in the Solar environment of gas and dust. Different scenarios are possible \citep[e.g., Grand Tack as described by][]{Morbidelli2011}. 
We know that in our Solar System no stable terrestrial planet could exist between
Jupiter and Saturn -- they all formed inside the Jovian zone.

We looked into the literature to find an interesting system quite similar to our own and found that HD~141399 would be a good
example \citep{Vogt2014}. The HZ in HD~141399 is located between two gas giants 
(see Table~\ref{Table2}), namely between planet {\bf c} (a = 0.7~au) and planet {\bf d} (a = 2.1~au). Calculation of the boundary values for the habitable zone shows that it lies between 0.8~au and 2.0 ~au from the star \citep{Kasting2007}. 

HD~141399 is a K0V star with m~=~1.07~$\mathrm{m_{Sun}}$, r~=~1.46~$\mathrm{r_{Sun}}$, and $\mathrm{T_{eff}}$~=~5600~K. The system was presented
by \cite{Vogt2014} who used radial velocity data sets from Keck-HIRES and the Lick Observatory's
Automated Planet Finder Telescope and Levy Spectrometer on Mt. Hamilton. 91 observations 
from over 10.5~yrs were analyzed and the parameters of four planets shown in
Table~\ref{Table2} have been found. 
Interestingly the planets HD~141399~b and HD~141399~c are close to the 2:1
mean motion resonance (MMR) which might give a clue on the history of the
system. \cite{Vogt2014} argue, that a 
possible inward migration would point towards a capture in resonance, which cannot be the case, given the values derived from the observations. 
Conversely, \cite{Batygin2013} show, that a divergence away from exact
resonance behavior is a natural outcome of dissipative evolution of resonant planetary pairs.
\cite{Vogt2014} carefully describe the method of determining the orbital
elements of the planets and checked their stability, which seems to be
guaranteed because of the almost circular orbits of the four planets.

\begin{table}
\begin{center}

\begin{tabular}{c|cccc}
Planets & Period (d) & e & a (au) & mass ($\mathrm{m_{Jupiter}}$) \\
\hline
HD 141399 b & 94  & 0.04 & 0.4225 & 0.46 \\
HD 141399 c & 202 & 0.05 & 0.7023 & 1 36 \\
\hline
3:2 MMR & 732 &  & 1.63& \\
\hline
HD 141399 d & 1070& 0.06 & 2.1348 & 1.22 \\
HD 141399 e & 3717& 0.00 & 4.8968 & 0.69 \\
\hline
\end{tabular}
\caption{Properties of the planets in the system HD~141399. The 3:2 MMR is called "Hilda region" and resides in the HZ of HD~141399.} 
\label{Table2}
\end{center}
\end{table}

\section{The Method}\label{sec:method}

The method used for the investigation of the formation of possible terrestrial planets between HD~141399~c and HD~141399~d was the Lie-ingration method
which has already been used extensively by our group \citep[e.g.,][]{Dvorak1986, Delva1984, Hanslmeier1984, Lichtenegger1984}. It is well adapted for the integration of the equations of motion for planetary systems hosting massless and massive bodies and treats close encounters and collisions with a
high precision due to its automatically chosen step size. The method has been compared to
other methods in detail by \cite{Eggl2010}. 

We ask the question where and how these terrestrial planets could have formed in
HD~141399. For being habitable the planets need to have water and therefore we
included water content of the planetesimals in our formation computations.

\begin{figure*}                                                                                                                               
\begin{center}                                                                                                                               
\includegraphics[angle=270, width=0.45\textwidth]{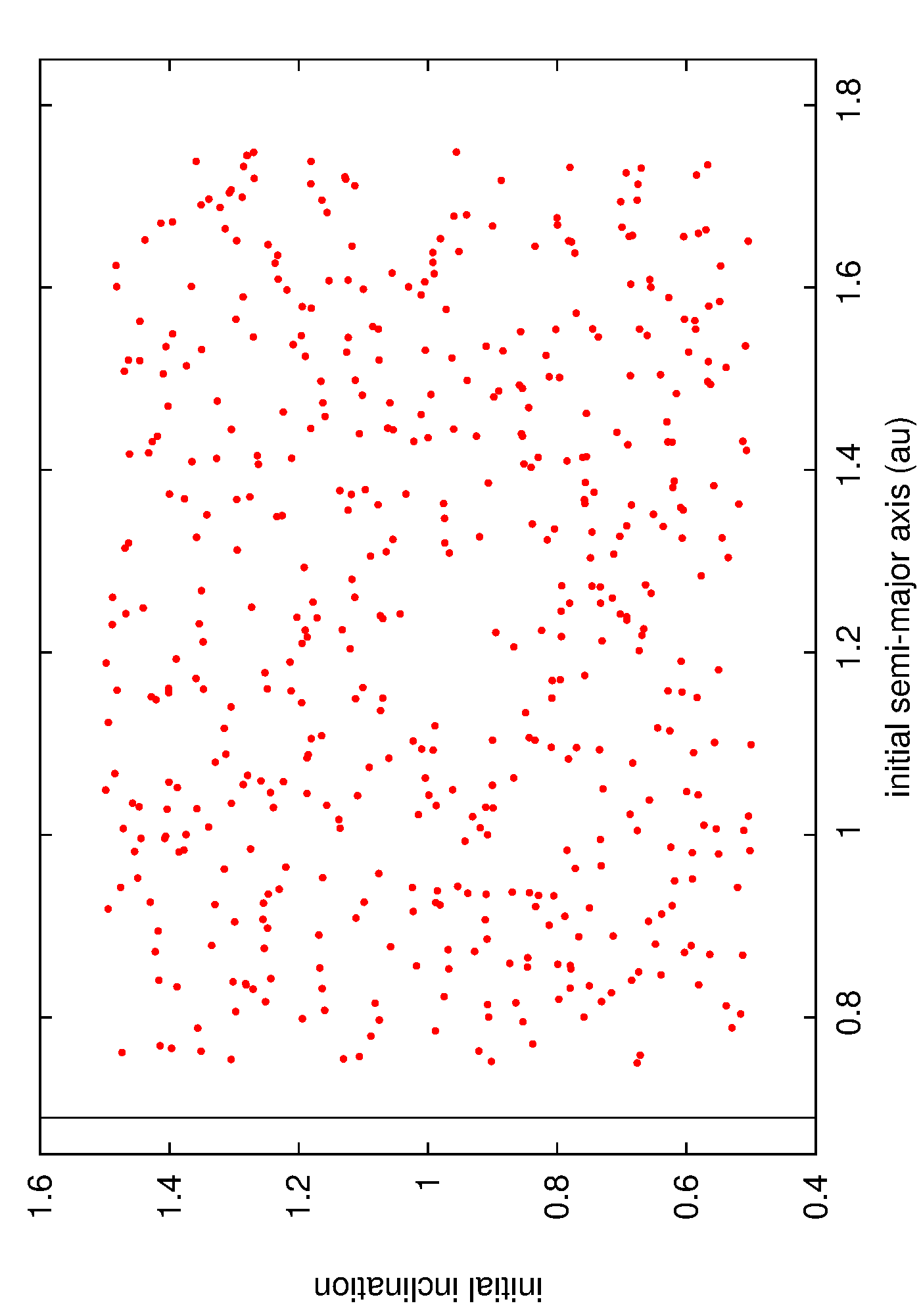}
\includegraphics[angle=270, width=0.45\textwidth]{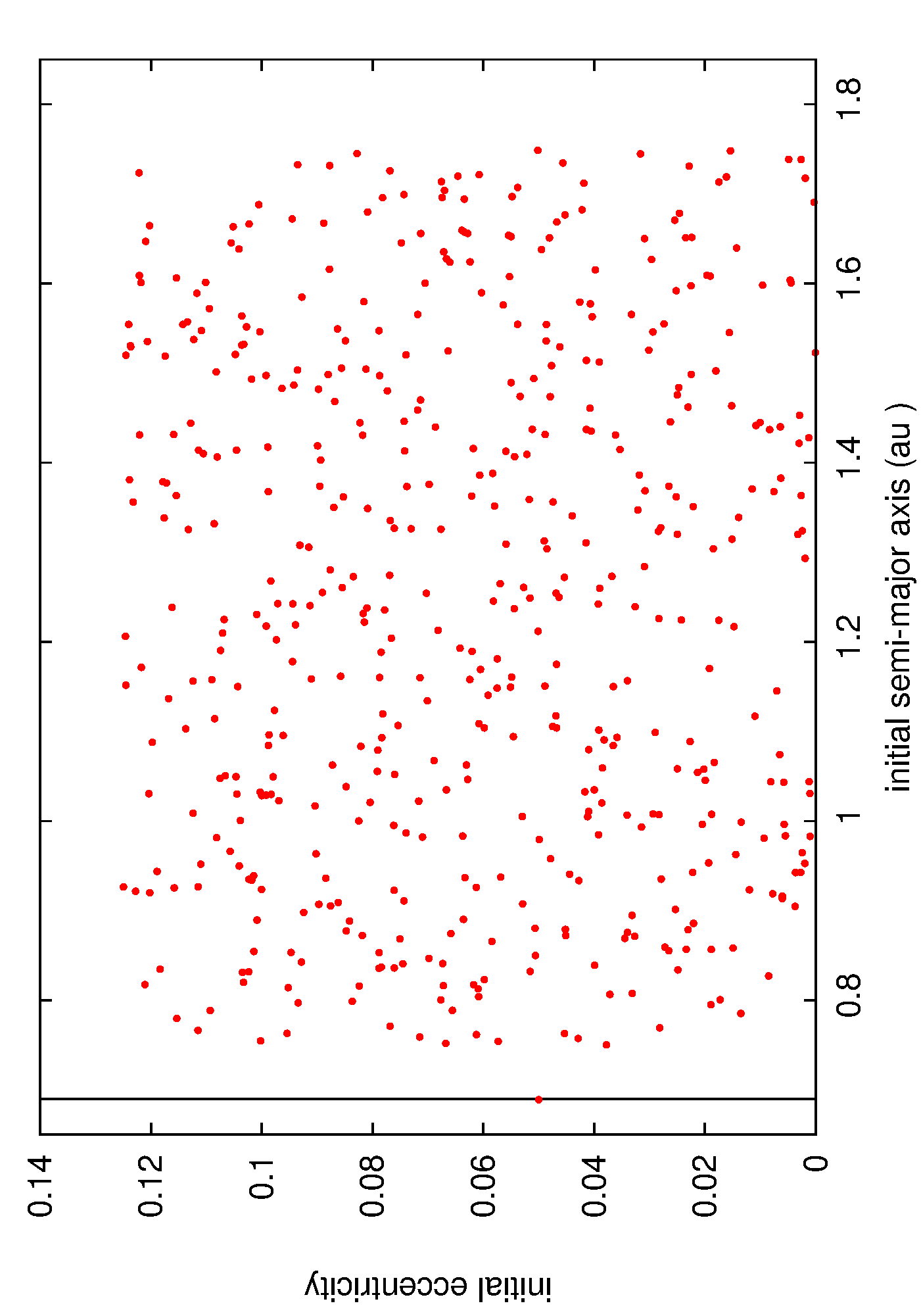}
\caption{Example of the initial inclinations and eccentricities versus semi-major axis. The orbital elements were chosen randomly for each of the different runs. The vertical line marks the position of planet {\bf c}.}\label{Fig.ini}                                                                                                                                
\end{center}                                                                                                                               
\end{figure*}

Making use of the Lie integration method four main runs were undertaken:

\begin{itemize}

\item First we distributed 500 massless bodies in the region between 0.7~au
and 2.1~au (the location of planet {\bf c} and planet {\bf d}) with initial
conditions randomly distributed for the orbital elements: semi-major axis a ($0.75 \, au < \mathrm{a} < 1.75 \, au$), eccentricity $\mathrm{e}  < 0.13$, inclination i ($0.5^{\circ} < \mathrm{i} < 1.5^{\circ}$), $\Omega = 154^{\circ}$, and $\omega = 0.5^{\circ}$. The mean
anomaly was randomly distributed. The whole system was integrated for 1~Myr for 8 differently chosen initial conditions. See Figure~\ref{Fig.ini}.

\item In the next step we replaced the massless bodies with massive ones ($0.92 \, \mathrm{m_{Moon}} < m < 1.23 \, \mathrm{m_{Moon}}$) with a total mass of approximately 5 $\mathrm{m_{Earth}}$, which was for example also used in \cite{Quintana2014}\footnote{several authors used initially different values of up to 10 Earth-masses \citep[e.g.,][]{Raymond2006}.}. The orbital elements were initially chosen randomly like in the first run. These computations were performed for a total set of 50 different initial conditions.

\item Again the initial conditions were chosen randomly but we took into account
that the estimated masses by the observers are only spectroscopically determined. Thus
three additional computations were performed for massless bodies but with more massive four giant planets \citep[the most probable masses given by the most probable inclinations,][]{Tscharnuter1984}.

\item In a forth run we checked the outer region between planets {\bf d} and {\bf e} by populating it with massless bodies (see Figure~\ref{Fig.2}). 

\end{itemize}

The water content of the planetesimals was calculated to be initially between 5\% to 10\% according to the mass of the body and the distance to the central star.

We note that the number of bodies diminished with integration time because of 

\begin{itemize} 

\item collisions amongst them resulting in formation of bigger planets (for the computations with massive planetesimals -- perfect merging was assumed\footnote{the parameters of each collision (impact velocities, impact angle) have been stored for possible follow up studies with SPH (Smoothed Particle Hydrodynamics).})

\item collisions with one of the gas giants

\item collisions with the host star 

\item escape from the system after a close encounter with one of the giant planets (as escape criterion for a body we fixed: e > 0.99). 
\end{itemize} 

\section{The 'Belt' of Massless Bodies}\label{sec:belt}

The reason for the choice of a close investigation of the region between
planets {\bf c} and planet {\bf d}  (we call it the belt) is the following: we are interested in the
formation of possible terrestrial planets in the HZ of HD~141399. 

In a recent work of \cite{Agnew2018} amongst other multi-planetary systems with gas giants the dynamical stability of HD~141399 is proven, using best-fit planetary and stellar parameters from the NASA Exoplanet Archive. They show that a potential terrestrial planet (m $\sim$ 1 $\mathrm{m_{Earth}}$) can inhabit a stable orbit inside the HZ of this particular system for the integration period of $10^7$ yrs.

In order to have a first overview of the dynamics of orbits inside this belt we
compute 8 'packages' of 500 massless bodies each. All the orbits were integrated for 1 Myr. 

\subsection{MMR Induced Structures in the Belt}

\begin{figure*}                                                                                                                               
\begin{center}                                                                                                                               
\includegraphics[width=0.75\textwidth]{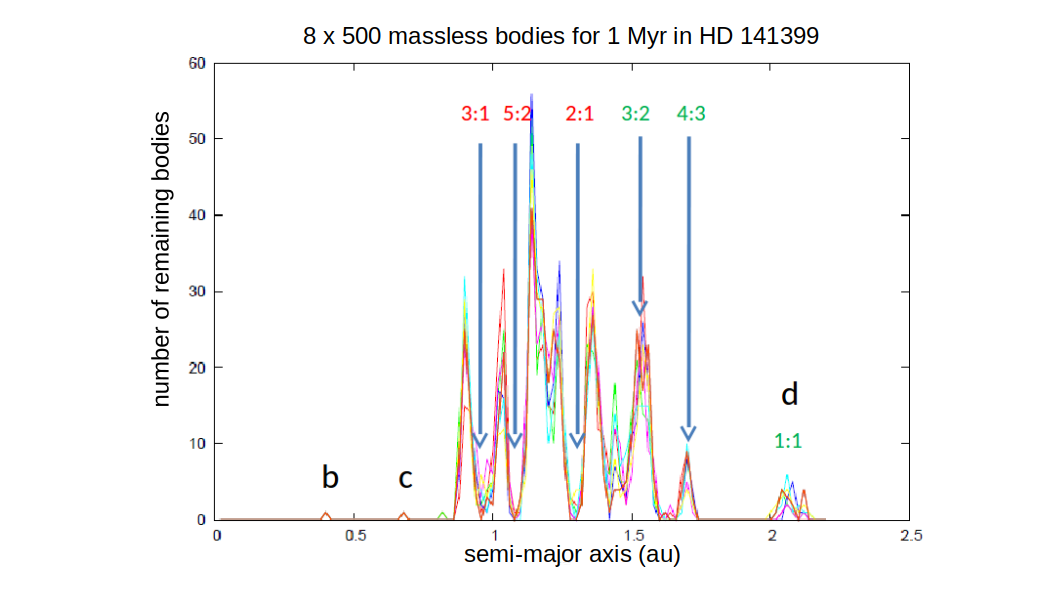}
\caption{The results of the integration of 8 x 500 massless bodies
between the planets {\bf c} and {\bf d}. The number of remaining bodies is plotted separately for each of the 8 runs (y-axis) versus the semi-major axis (x-axis). The gaps and 'groups' are very well visible. The numbers describe the resonances with planet \textbf{d}. Note that the green numbers show stabilizing resonances, while the red numbers show destabilizing resonances.}\label{Fig.1}                                                                                                                                

\end{center}                                                                                                                               
\end{figure*}

The structure visible in Figures~\ref{Fig.1} and \ref{Fig.10} looks very similar to the Kirkwood gaps in the main belt of asteroids in our Solar System. In Figure~\ref{Fig.1} we marked the MMRs.

\begin{figure}                                                                                                                               
\begin{center}                                                                                                                               
\includegraphics[angle=270,width=8.4cm]{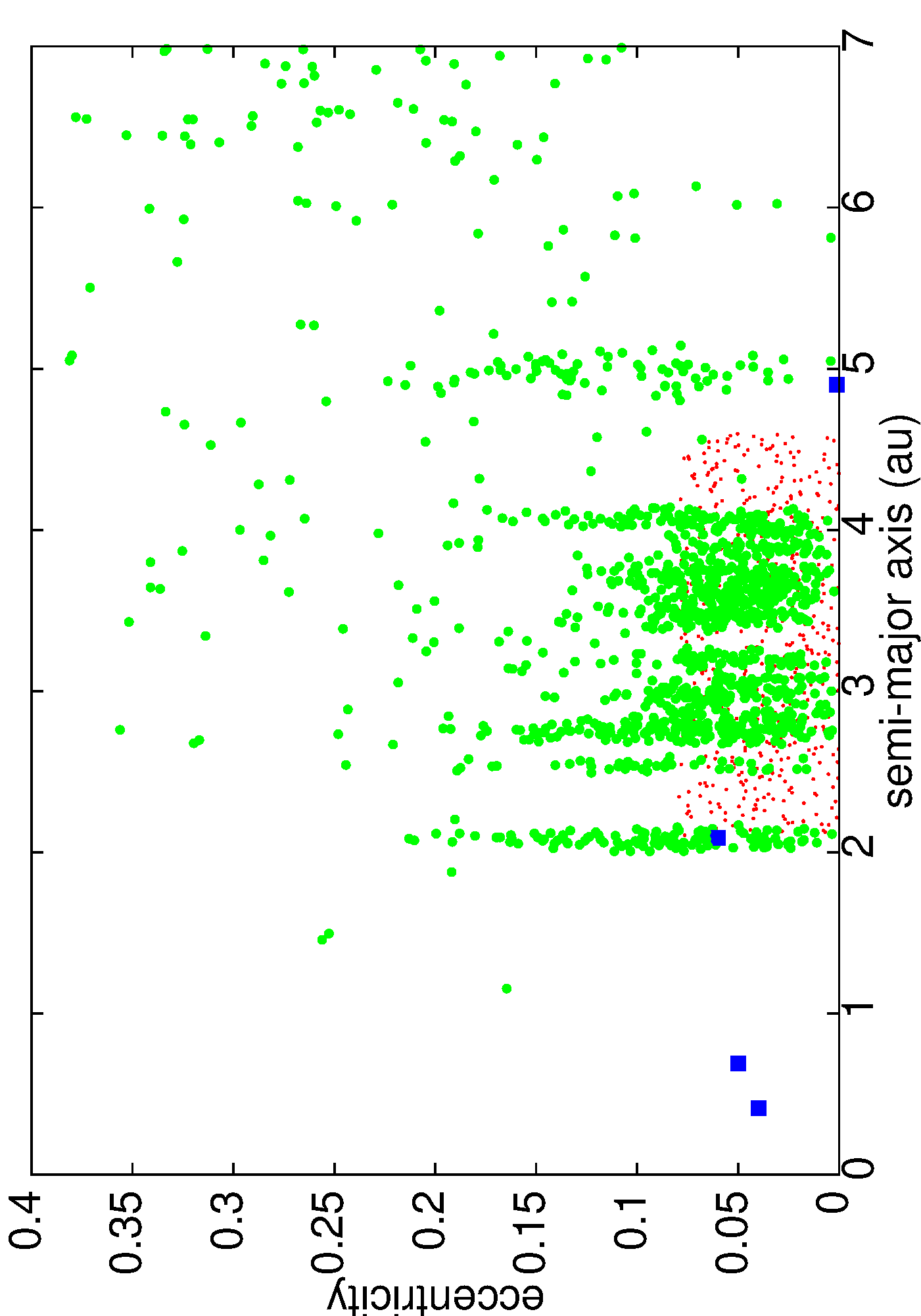}                                                                         
\caption{Results of the integration of 3 x 500  massless bodies
(with slightly different initial conditions between the planets
d and planet e). Semi-major axis (x-axis) versus eccentricity (y-axis). The number of remaining bodies is plotted in green, the initial distribution in red and the 4 planets as blue squares. Also visible are gaps and groups as for the inner belt.}\label{Fig.2}                                                                                                                                
\end{center}                                                                                                                                 
\end{figure}

When we compare the results of an integration of massless bodies between the planets {\bf d} and {\bf e} in Figure~\ref{Fig.2} with the region between Jupiter and Saturn in our Solar System there is a big difference evident. In the Solar System there are no stable families of asteroids with small eccentricities between $5.2\, au < a < 9.52\, au$. In HD~141399 we found accumulations of bodies in certain resonances. Even in the equilateral Lagrange points of planet {\bf d} and planet {\bf e} many planetesimals are captured and form Trojan families. 

The extension of computing bodies outside of planet {\bf d} has been undertaken for the sake of completeness of the dynamical study of HD~141599. It becomes relevant as one recognizes the Trojans around planet {\bf d} (see Figure~\ref{Fig.2}). This was one of the reasons to explore the Trojan region of planet {\bf d} in more detail (see Appendix). Note that around Saturn there is no a Trojan cloud \citep{Dvorak2008, Dvorak2014}. Nevertheless, it would be worth to do a more detailed investigation.

\subsection{Larger Masses}\label{sec:largemass}

\begin{figure}                                                                                                                               
\begin{center}                                                                                                                               
\includegraphics[angle=270,width=8.4cm]{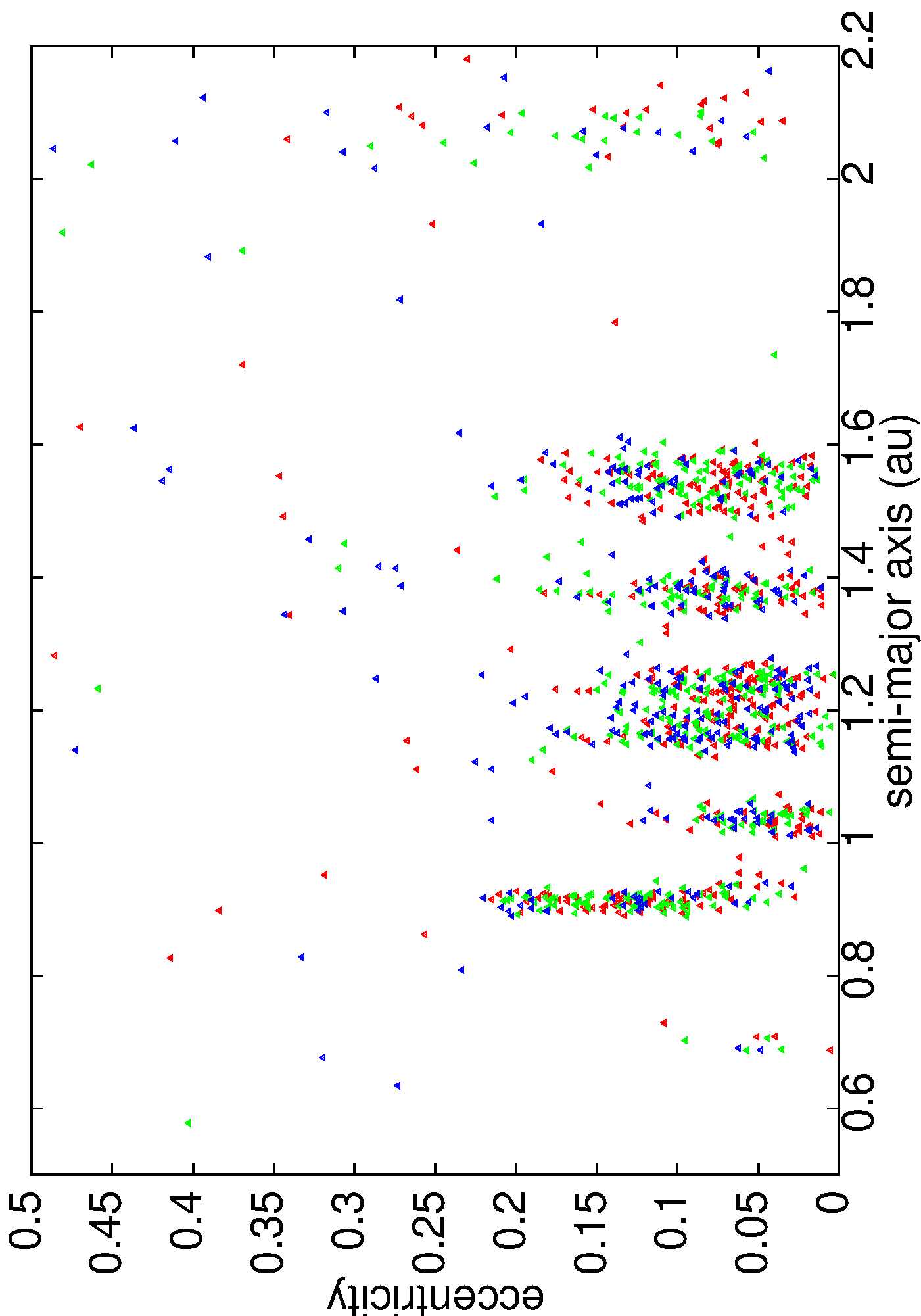}   
\caption{Belt of massless bodies for increased masses of the giant planets {\bf c} and {\bf d} (as described in Section~\ref{sec:largemass}). Semi-major axis is plotted versus eccentricity with different colors indicating different runs.}\label{Fig.10}                                                                                                                                

\end{center}                                                                                                                              
\end{figure}

The following Section deals with a check of stability and formation of families of massless bodies in the system of HD~141399 assuming larger masses for the four gas giants. We performed these computations with three different initial conditions (compare Section~\ref{sec:method} and Figure~\ref{Fig.10}).

For these computations we assumed the most probable inclination of the orbital
plane of the giant planets  -- the statistical mean value. We assume equally distributed inclinations \citep[$\overline{\sin^3(i)}=0.56$,][]{Tscharnuter1984} which leads to an increasing factor of 0.59 and gives $\mathrm{m_{planet\,b}}$~=~ 8.035, $\mathrm{m_{planet\,c}}$~=~2.378, $\mathrm{m_{planet\,d}}$~=~2.127 and $\mathrm{m_{planet\,e}}$~=~1.15 (masses in $\mathrm{M_{Jupiter}}$). Because of the small eccentricities, the gas giants are stable for Gyrs in this configuration, too. For the distribution of massless bodies in the HZ one can observe that gaps and groups appear as in our previous computations (see Figure~\ref{Fig.10}). 
A comparison of Figures~\ref{Fig.1} and \ref{Fig.10} shows similarities of the gaps and groups (areas of accumulation of bodies) in the distribution of bodies after an integration time of 1 Myr. These gaps and groups result from the acting of resonances (MMRs) with the giant planets. 

\section{Some Examples of New Terrestrial Planets}\label{sec:examples}

\begin{figure}                                                                                                                               
\begin{center}                                                                                                                               
\includegraphics[width=6.4cm,angle=270]{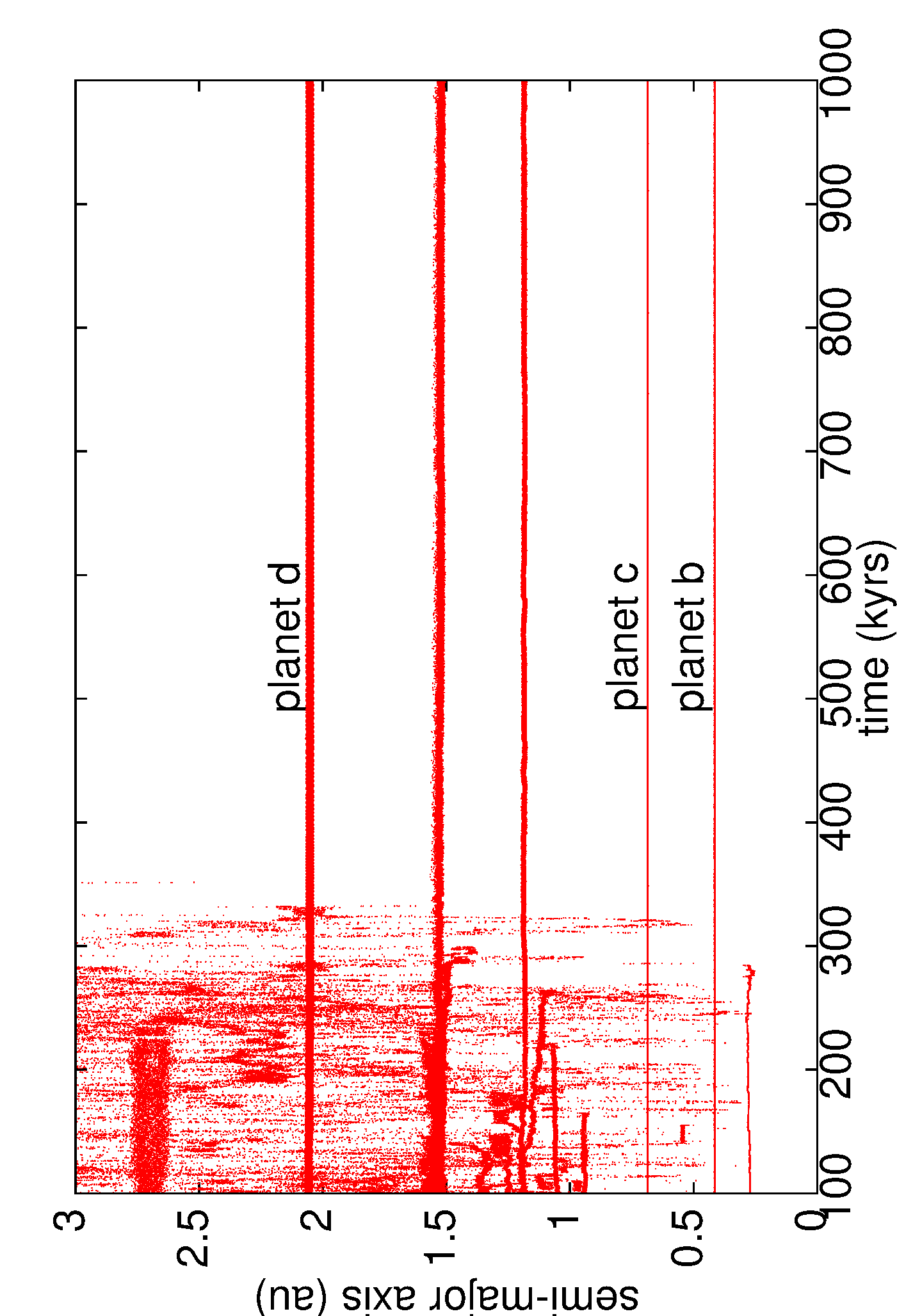}              
\caption{Orbits of bodies during 0.1 and 1 Myr. Time is plotted versus semi-major axis. Many bodies escape between 0.1 and 0.35 Myr. Two terrestrial planets survive between planet {\bf c} and planet {\bf d}. The smaller one at 1.5~au has approximately Mars-mass while the bigger one at 1.1~au has approximately 0.5 Earth-masses. The bottom line shows a small body with a high eccentricity. Compare Figure~\ref{Fig.6} which shows results of the same run.}\label{Fig.5}                                                                                                                                
\end{center}                                                                                                                                 
\end{figure}

As an example of the formation process taking place in the system of HD~141399 we picked out one simulation for discussion. Semi-major axis versus mass of the bodies of this particular simulation are shown in Figure~\ref{Fig.5}. One can observe that after 0.1 Myr most of the original Moon-sized objects are gone either due to collisions (with one of the gas giants or the host star) or they were thrown out of the system after close encounter(s) with one of the gas giants. A close encounter can lead to large eccentricity (e $>$ 0.7) and subsequent ejection from the system. 
In the presented simulation two massive planets were formed -- one with the mass of Mars and one with approximately 0.5 Earth-masses. They stay in stable orbits between planet {\bf c} and
planet {\bf d} for 1 Myr. An extension of the integration time to 10 Myr in this particular case shows that they obtain stable orbits over the whole integration time.

\begin{figure}                                                                                                                               
\begin{center}                                                                                                                               
\includegraphics[width=6.4cm, angle=270]{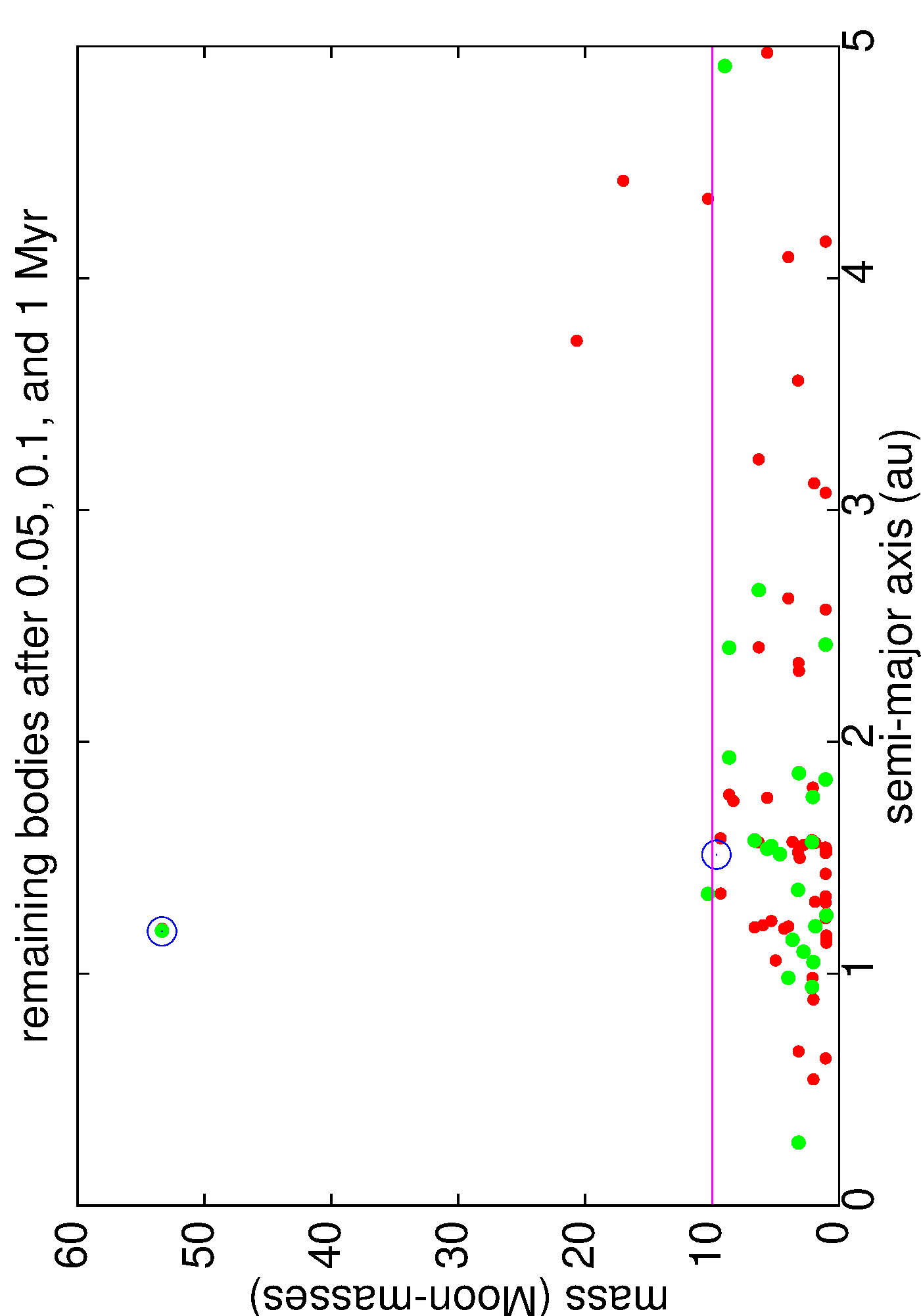}   
\caption{Remaining bodies after 0.05 Myr (red points), 0.1 Myr (green points), and 1 
Myr (open blue circles). Semi-major axis is plotted versus mass of the bodies in ($\mathrm{m_{Moon}}$). The horizontal line marks the size of 10 Moon-masses for clarity. The orbits of the planets depicted by the two open blue circles are shown in Figure~\ref{Fig.5}.}\label{Fig.6}                                                                                                                                
\end{center}                                                                                                                              
\end{figure}

In Figure~\ref{Fig.6} mass versus semi-major axis of the remaining bodies after 0.05 Myr, 0.1 Myr, and 1 Myr are shown. We remark that after 0.05 Myr (red points) there are still many -- now protoplanets -- present in the system. Additionally still some of the
originally Moon-sized bodies remain. We observe the formation of a planet of 20
Moon-masses at $\sim$3.8 au, another one with 16 Moon-masses at $\sim$4.4 au, and one with 10 Moon-masses at $\sim$4.3~au within the first 0.05~Myr. 

Due to their eccentricities, which allow close encounters with planet {\bf d} and subsequent collision or ejection (not checked in detail here) the number of remaining bodies after 0.1 Myr (green points) is small. 

Note that Figure \ref{Fig.5} and Figure \ref{Fig.6} show results of the same simulation while Figure~\ref{Fig.7} depicts results from another run.

\section{The Growing from Planetesimals to Protoplanets}\label{sec:growth}

There are two ways of looking at the the growth of planets: 
\begin{itemize}
\item[a)] follow one particular body during its evolution (monitoring each collision in order to watch the growth to the final
protoplanet/ planet)
\item[b)] follow the development of the whole belt of bodies (checking the number of remaining bodies and the growth after several time steps)
\end{itemize}

\begin{figure}                                                                                                                               
\begin{center}                                                                                                                               
\includegraphics[width=8.4cm]{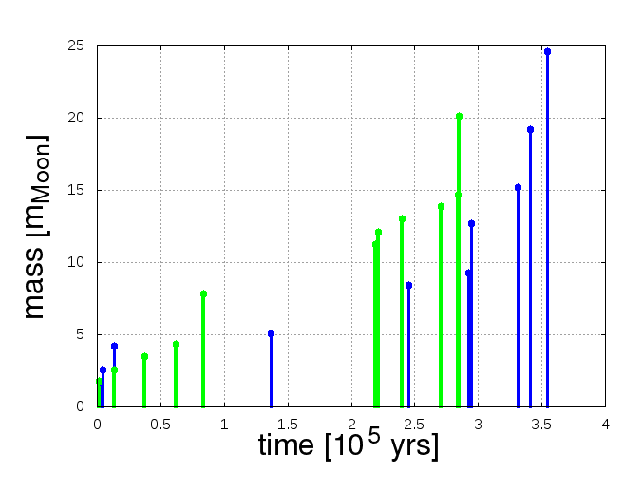} 
\caption{Growth process of two terrestrial planets. The blue lines show the formation process of the most massive planet formed in this particular simulation. The green lines depict the growth of the second most massive planet formed during the integration. The axes show time versus mass.}\label{Fig.7}                                                                                                                                

\end{center}                                                                                                                              
\end{figure}

We follow the growth of a planetesimal to the moment of reaching the mass of a terrestrial planet.
A special case is shown in Figure~\ref{Fig.7} where after 9 collisions (perfect merging was assumed) a terrestrial planet is formed. The body considered has a mass of 25 $\mathrm{m_{Moon}}$ (blue lines in Figure~\ref{Fig.7}). The second most massive one has only 20 $\mathrm{m_{Moon}}$ (green lines in Figure~\ref{Fig.7}). 

The growth process can be divided into different steps: In the beginning small bodies collide and form larger and larger objects until they reach a mass of several Moon-masses. The collision of two bigger bodies leads to the formation of a terrestrial planet like object. Yet, there are still many small bodies around which are 'added' to the big body via collisions.

\subsection{Example of the Evolution of the Belt}

\begin{figure*}
\begin{center}
\includegraphics[width=0.49\textwidth]{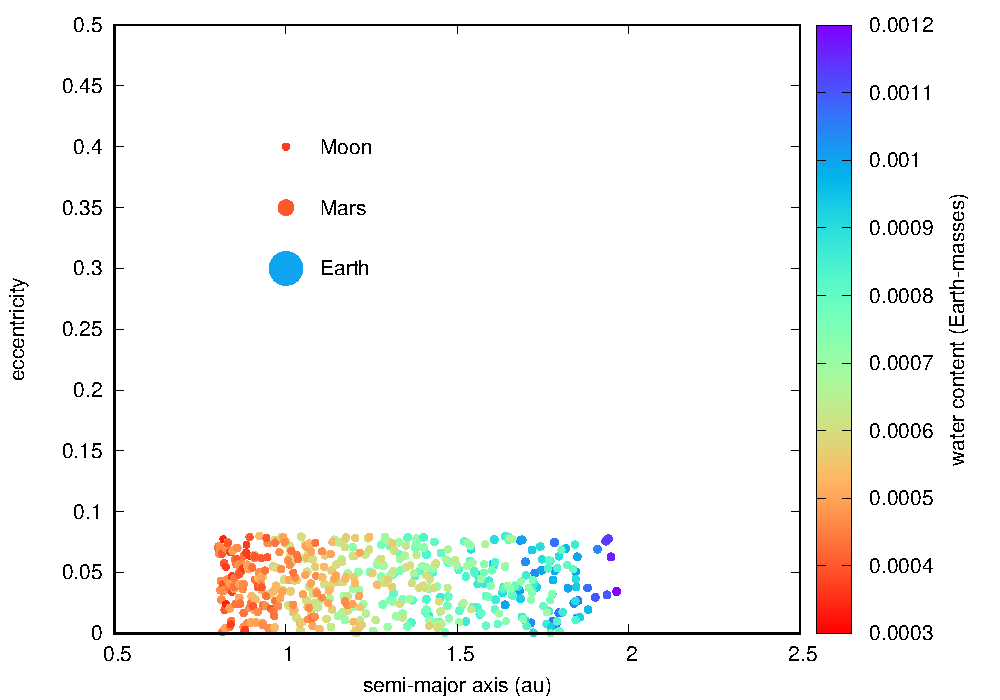}
\includegraphics[width=0.49\textwidth]{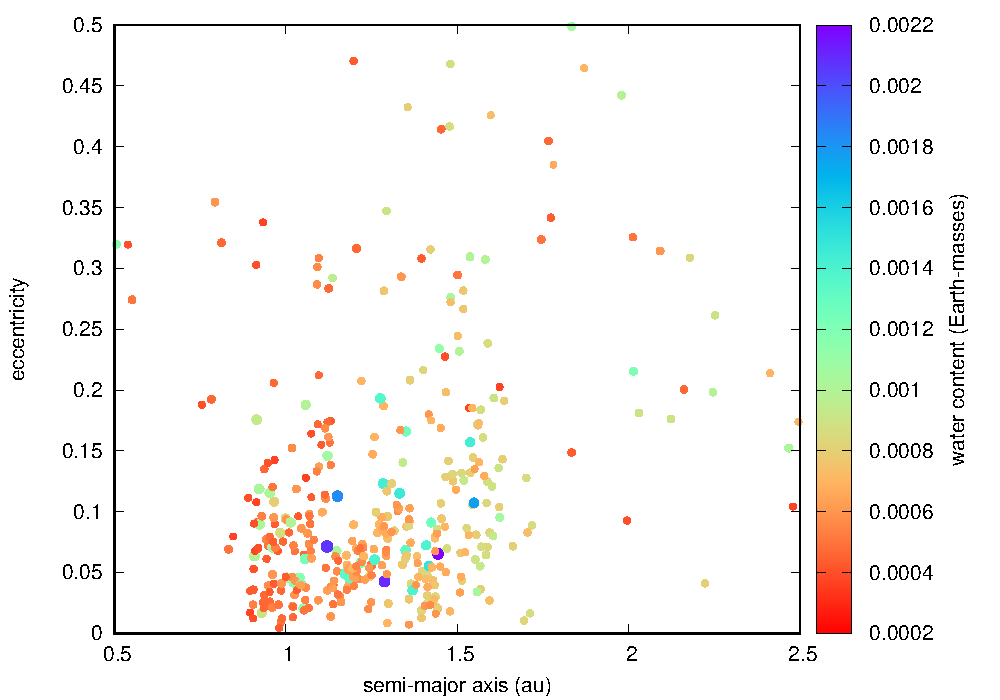} \\
\includegraphics[width=0.49\textwidth]{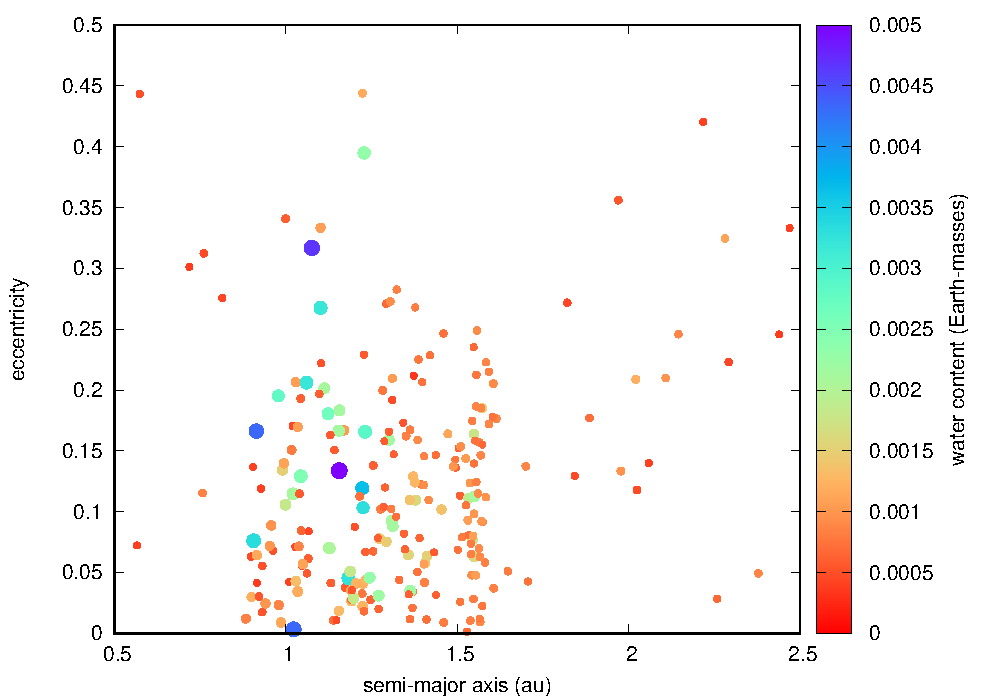}
\includegraphics[width=0.49\textwidth]{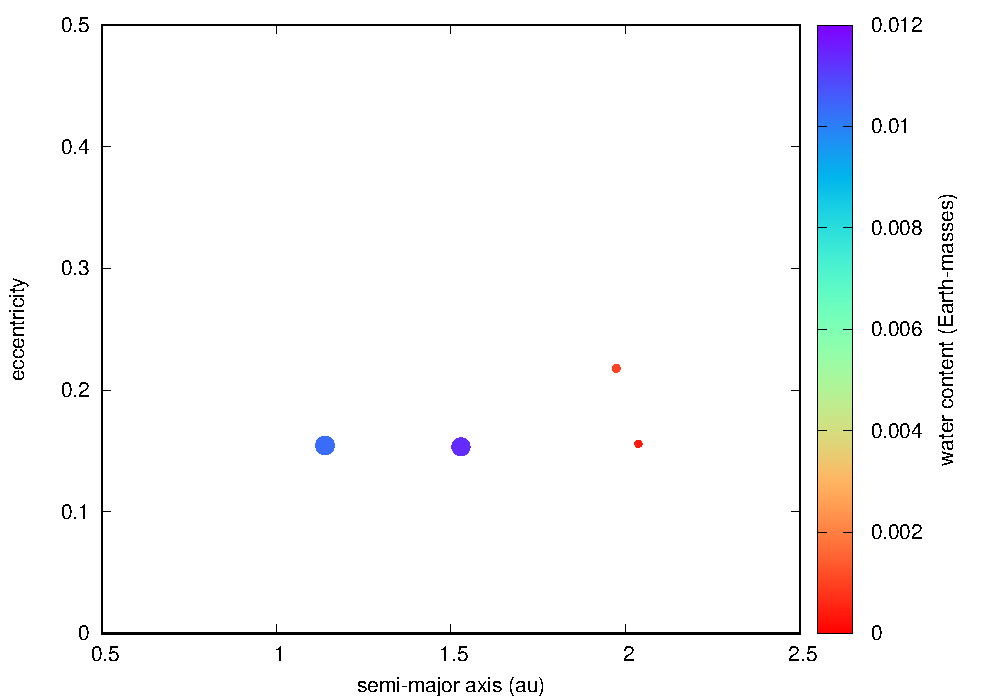}
\caption{ One simulation out of 50: Initial conditions (upper left). Remaining bodies after 1 kyr (upper right), 10 kyr (lower left), and 1 Myr (lower right). Semi-major axis is plotted versus eccentricity. The size of the bodies is proportional to the size of the dots (for comparison Earth-, Mars-, and Moon-mass are shown in the upper left panel). The water content of each body is represented by its color.} \label{Fig.8}                                                                                                                                                                                                           
\end{center}                                                                                                                                 
\end{figure*}

We show one out of 50 integrations and follow the 500 Moon-sized bodies over time in order to depict the evolution of the belt. We show the initial conditions and snapshots after 1 kyr, 10 kyr, and 1 Myr in Figure~\ref{Fig.8}.

One can observe that the eccentricities grow quite fast when we compare in
Figure~\ref{Fig.8} the initial distribution to the distribution semi-major axis versus
eccentricity after 1000 years (left and right upper panels in Figure~\ref{Fig.8}). Bodies of the
size of several times the Moon are formed during such a short time. After
10 kyrs  we can see that already some Mars-sized protoplanets were formed;
like in the former plot one recognizes also that the majority of bodies is now
confined primarily between 0.9~au and 1.5~au (lower left panel in Figure~\ref{Fig.8}).These larger
bodies disappear during the continued integration of the equations of motion
up to 1 Myrs; we can explain it by encounters with the gas giants which throw
them far away from the central star to large semi-major axes and eccentricities
and may also lead to escapes.
But even collisions with the four large planets let them 'disappear'.
The left over after 1 Myr consists now of only two Mars-sized
planets and two Moon-like ones (lower right panel in Figure~\ref{Fig.8}).

\section{Main Results for the Formation of Terrestrial Planets}\label{sec:results}

The results from the integrations are summarized in the following plots which show (note that the water content will be discussed in Section~\ref{watercontent})

\begin{figure}                                                                                                                               
\begin{center}                                                                                                                               
\includegraphics[width=8.4cm]{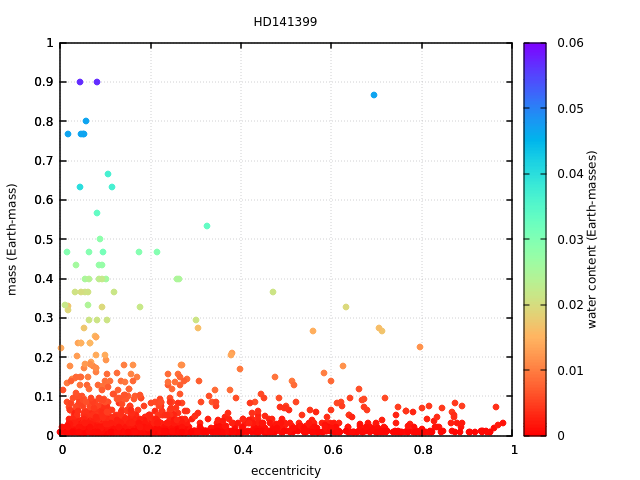}
\caption{Accumulated bodies of 50 different runs with initially 500 Moon-sized bodies between planet {\bf c} and planet {\bf d} after 1 Myr. Semi-major axis is plotted versus mass with the color indicating the water content (in Earth-masses) of each planet.}\label{Fig.15b}                                                                                                                                
\end{center}                                                                                                                                 
\end{figure}

\begin{figure}                                                                                                                               
\begin{center}                                                                                                                               
\includegraphics[width=8.4cm]{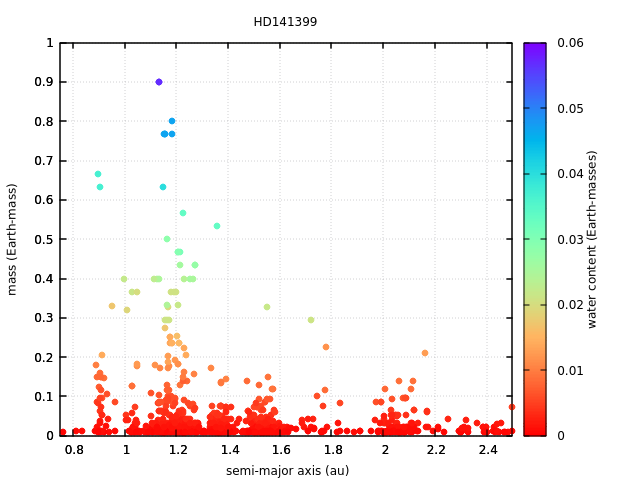}
\caption{Same as Figure~\ref{Fig.15b} with the color indicating the water content (in Earth-masses) of each planet.}\label{Fig.15a}                                                                                                                                
\end{center}                                                                                                                                 
\end{figure}

\begin{figure}                                                                                                                               
\begin{center}                                                                                                                               
\includegraphics[width=8.4cm]{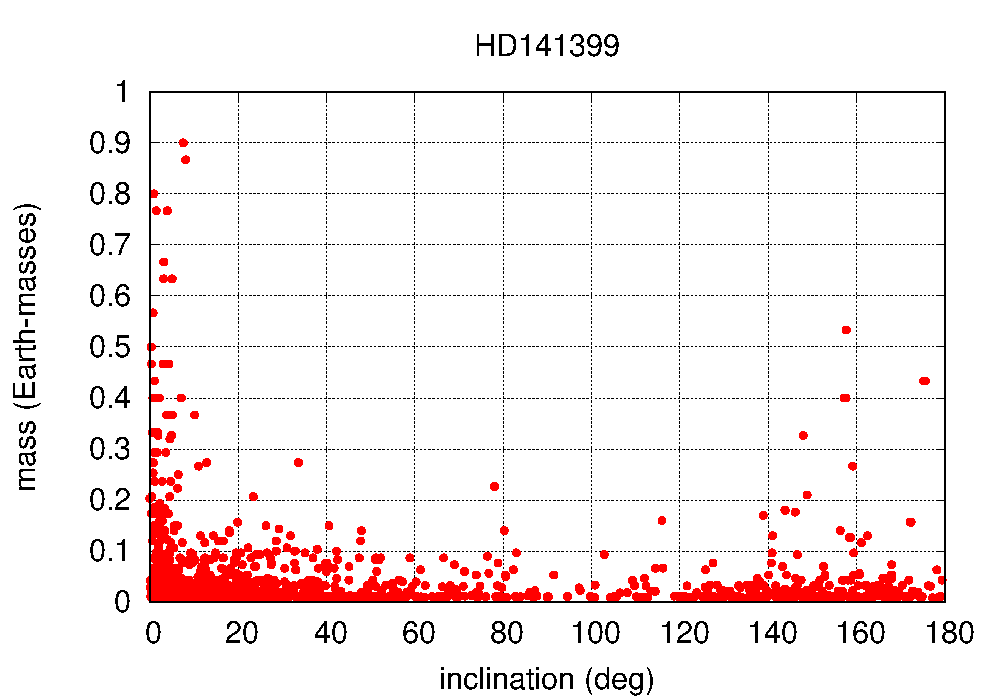}
\caption{Same as Figure~\ref{Fig.15a} but for mass versus inclination.}\label{Fig.16}                                                                                                                                
\end{center}                                                                                                                                 
\end{figure}

\begin{figure}                                                                                                                               
\begin{center}                                                                                                                               
\includegraphics[width=8.4cm]{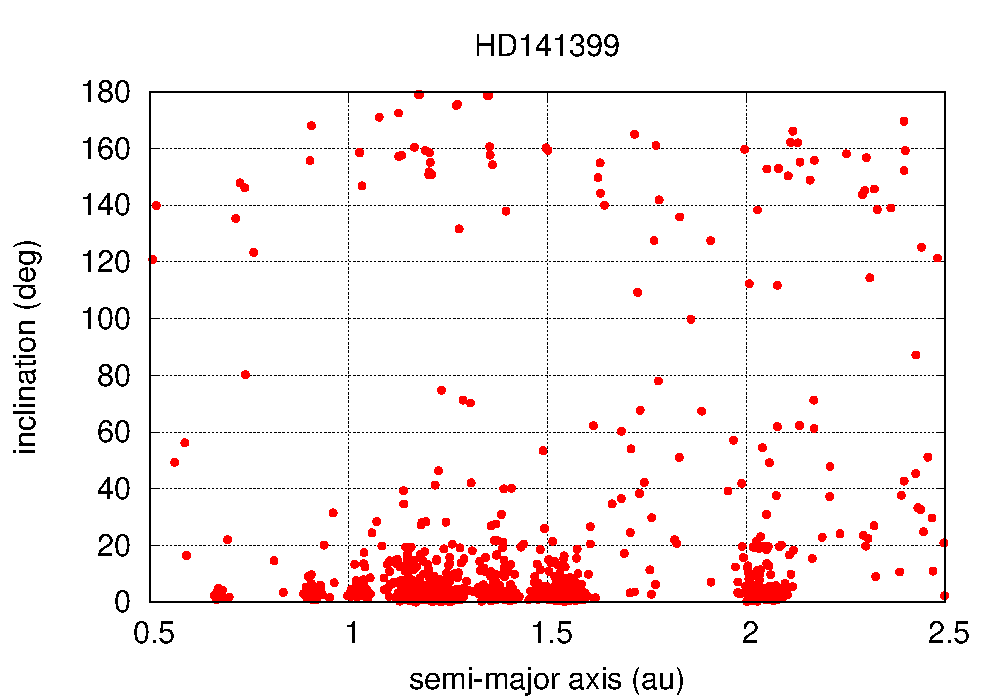}
\caption{Same as Figure~\ref{Fig.15a} but for semi-major axis versus inclination. One can see that some of the bodies captured as Trojans of planet {\bf d} have inclinations of up to $80^{\circ}$.}\label{Fig.inclination}                                                                                                                                
\end{center}                                                                                                                                 
\end{figure}

\begin{figure}                                                                                                                               
\begin{center}                                                                                                                               
\includegraphics[width=8.4cm]{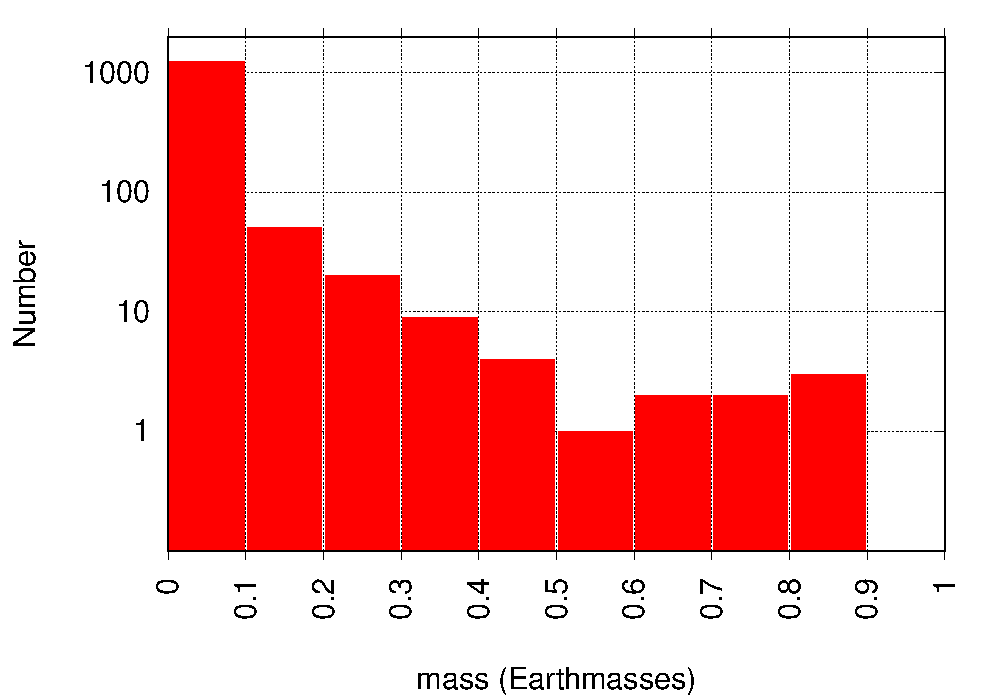}
\caption{Histogram of the planets depicted in Figure~\ref{Fig.15a}. Number versus mass. One can see that most of the planetesimals remained small for the whole integration time. Nevertheless some were able to gain a mass of up to 0.9 Earth-masses.}\label{Fig.17}                                                                                                                                
\end{center}                                                                                                                                 
\end{figure}

\begin{itemize}

\item Figure~\ref{Fig.15b}: We show the resulting planets formed out of 50 x 500 planetesimals distributed between 0.7~au and 2.1~au (between planet {\bf c} and {\bf d}). Only two of the planets with mass comparable to the mass of Earth were formed via collisions and consecutive merging (these planets gained 0.9 Earth-masses). In addition many Mars-sized planets are formed. One can observe that most of the more massive planets inhabit low eccentric orbits. The biggest bodies have eccentricities e $<$ 0.1, which assures that they are in relatively stable orbits. Nevertheless, there are some bodies with mass of $\sim$ 0.5 Earth-masses with eccentricities up to $e \sim 0.3$ which might experience close encounters or even collision with one of the gas giants. Eventually they might fall into the star or even be ejected from the system.

\item Figure~\ref{Fig.15a}: The distribution of bodies at the end of the integration (0.5 Myr) shows that most of the bodies remain between the planets \textbf{c} and \textbf{d}. An accumulation of objects can be observed at certain semi-major axes. The majority of the formed planets with up to 0.9 Earth-masses reside in the area $0.8 \, \textrm{au} < a < 1.6 \, \textrm{au}$.
Some objects accumulate around the gas giant {\bf d}. They have been captured into the 1:1 MMR as Trojan planets. Furthermore there appears to be a more or less empty region inside of planet \textbf{d}.
Clearly, empty regions analogue to the 'Kirkwood gaps' in our Solar System are seen. The groups where many larger planets are formed, are observable at 1.15~au (by far the biggest one), 1.35~au, and 1.5~au.

\item Figure~\ref{Fig.16}: The distribution of bodies at the end of the integration shows that most of them stay in the plane of the gas giants. In particular the more massive ones show only small values for inclination (i $<$ $10^{\circ}$). In some simulations bodies were scattered to retrograde orbits during the formation process. 

\item Figure~\ref{Fig.inclination}: The plot shows, that some of the bodies captured as Trojans of planet {\bf d} have orbits with inclinations of up to $80^{\circ}$. Some are even on retrograde orbits. Nevertheless, the majority of bodies stays in the plane of the gas giants with $i < 20^{\circ}$.

\item Figure~\ref{Fig.17}: Statistical summary of all the remaining bodies after the end of the integration (included are all 50 runs). One can see that most of the planetesimals remained small. Nevertheless, some (to be exact: 2) managed to gain a mass of up to 0.9 Earth-masses.

\end{itemize}

\subsection{The Role of Water Content}\label{watercontent}

At the beginning of the integration each of the approximately Moon-sized bodies in the belt between the planets {\bf c} and {\bf d} is assigned with a water mass fraction of 5 -- 10 \% depending on the distance from the star. We assume that after each collision the bodies perfectly merge. Figure~\ref{Fig.15b} and \ref{Fig.15a} show the summary of all 50 runs. Depicted are the mass and water content of the bodies after the end of the integration time. One can see that, obviously, the more massive ones collected more water (in units of Earth-masses). Nevertheless, the water content in percentages is the same for small and massive bodies. 

SPH (smooth particle hydrodynamics) computations of collisions between bodies show that they suffer from essential loss of water depending 
on the impact velocity, the impact angle etc. \citep{Burger2018, Maindl2013, Maindl2014, Sch2016}. All these parameters were not yet included in our
simulations. As we assume that no mass is lost during a collision we clearly overestimate the water mass fraction present in each planet at the end of the simulation.  

The more realistic models of collisions with the aid of SPH-codes
need time consuming simulations for every single collision and should be treated in combination with a n-body code. 

Nevertheless, our results show that the
planets formed in between the planets {\bf c} and {\bf d} could foster life as they remain on stable low eccentric orbits in the HZ around HD~141399 with possibly water on their surfaces.

\section{Conclusions}\label{sec:conclusion}

The goal of the presented work was to investigate the probability of the existence and formation of
terrestrial planets in an extrasolar planetary system with several gas giants.
We found a very interesting example: The system HD~141399, where a K0V star hosts 4 giant planets in distances to the central star between 0.4~au and 5.2~au. The small eccentricities of the planets guarantee stable orbits up to
Gyrs. This is true, even when it is taken into account that the determined masses via spectral analysis from observations with large telescopes are only minimum values.

In order to answer the question about existence and formation of terrestrial planets in the system of HD~141399 we assumed a leftover belt of hundreds of small bodies between the two planets {\bf c} and {\bf d} orbiting at 0.7~au respectively 2.1~au. Furthermore we assume that the formation of the four gas giants has been finished (see Section~\ref{sec:method}). The region between the planets {\bf c} and {\bf d} includes the habitable zone (HZ) where water on a rocky planet can be present in liquid form. This might give the planet the chance to develop and sustain life on its surface. 

The work was done in several steps. First of all we studied the evolution of orbits of
massless bodies in the HZ which yielded a structure similar to the 'Kirkwood Gaps' found in the main belt of asteroids in our Solar System -- all associated to mean motion resonances (MMRs) with the gas giants. 
These simulations have been done with the primarily determined masses of the four gas giants. In addition the integrations were repeated with masses of the four giants estimated for the statistical mean value of inclination. There is only a small difference in the final distribution of the remaining massless bodies inside the belt (comparison of Figure~\ref{Fig.1} and \ref{Fig.10}).

As a second step we integrated Moon-sized bodies for 500 different initial conditions between the planets {\bf c} and {\bf d} with randomly chosen orbital elements, small initial eccentricities ($e \sim 0.1$), and small
initial inclinations ($i \sim 1^{\circ}$). We provide numerical integrations
of the equations of motion in the Newtonian framework up to the formation of terrestrial planets. We carefully studied the collisions between these bodies under the assumption of complete merging. Being aware of this rough  estimation we started SPH (Smooth Particle Hydrodynamics) computations which will provide more realistic results for the collisions in a future project. We performed 50 different runs with different initial conditions for the 500 planetary embryos. In several plots we show the respective results -- e.g after how many collisions such a terrestrial planet is formed, how fast the formation happens in different stages of the evolution, how many
planets form, and in which distance to the host star most of the planets
are formed. 

We found that there is primarily one region around 1.15~au where many terrestrial planets formed within approximately 1~Myr (most of the massless bodies survived in this region in previous computations (see Figure~\ref{Fig.15a}). Also in two other regions around 1.35~au and 1.55~au planets formed. The latter one corresponds to the so-called Hilda region (named after the Hilda asteroids in the main belt
of asteroids in our Solar System) in the 2:3~MMR with planet {\bf d}. Those Hildas are less numerous and less massive. 

Concerning formation computations with larger masses for the gas giants we found that it is far more difficult to form such planets, because relatively soon the objects undergo perturbations by the giant planets leading to large eccentricities and subsequent close encounters and/or ejections. Our test computations with a system with larger masses of the primary bodies
has shown that already after some thousand years only half of the original planetesimals
survived. Although this number is comparable to the computation with the
nominal masses (see Figure~\ref{Fig.8}) there is a big difference insofar that for the
latter test 95 \% of the original bodies left the central region because of close encounters
respectively collisions with the gas giants and also the central star and not
even one bigger body was formed within this time (see Section~\ref{sec:largemass}).
  
The special case of a MMR with one of the giant planets -- the capture of a small body in 1:1 resonance -- a so called Trojan, was investigated separately. One can observe a few aggregated bodies up to the size of Mars in the Trojan region around planet {\bf d}. In the results of our integrations of fictitious massless Trojans for 1 Myr we found stable regions around planet {\bf d} up to an inclination of the Trojans of about $80^{\circ}$ (compare Figure~\ref{Fig.inclination}).

In our computations we also included the water content of the bodies and followed the collisional process of accumulation of more and more mass up to the final formation of a habitable terrestrial planet. In conclusion we can say that terrestrial like planets may exist in the extrasolar planetary system of HD~141399 and it might even harbor life.

\section*{Acknowledgements}

This research is supported by the Austrian Science Fund (FWF) trough grant S11603-N16 (B.L. and R.D. and L.Z.). The computational results presented have been achieved in part using the Vienna Scientific Cluster (VSC). This work has been supported by the National Natural Science Foundation of China (NSFC, Grants No. 11473016, and No. 11333002)
Special thanks to T. I. Maindl for critically reading the manuscript and his help with Gnuplot. 








\appendix

\section{Trojans Around Planet {\bf d}}\label{sec:trojans}

Inspired by the presented results on the accumulation of bodies around planet {\bf d} (shown in Figure~\ref{Fig.2}) and the importance of possible Trojan planets in extrasolar planetary systems \citep{Dvorak2008,Schwarz2007,Dvorak2004, Laughlin2002} we checked the stability of the Lagrange
point $L_4$: for a grid of initial conditions 100 by 60 (semi-major axis versus
initial inclination) we integrated the respective orbits of fictitious
massless Trojans for 1 Myr (Figure~\ref{fig:emax}). There is a big stable region around a = 2.1 au
visible between 1.95~au $<$ a $<$ 2.175~au (for the plane case). The horseshoe-like contour of the 
stable region extends up to an inclination of $\mathrm{i}=57^{\circ}$ . The stability
criterion was the libration in the angle  $\lambda$  
around the equilibrium point $60^{\circ}$ ahead
in the planets orbit on one side (not shown here) and the largest eccentricity achieved during the
integration. The almost black region shows that the libration point
itself is very stable; more to the edge the eccentricity reaches values up to
0.3 before they escape (yellow region). All the structures visible are
comparable to those found in other studies of Trojans in the
Solar system \citep[e.g.,][]{Dvorak2012}.

\subsection{Dynamical Map}\label{sec:dynmap}

Beside the maximum eccentricity map, we also construct the dynamical map of the Trojan region using the \textit{spectral number} (SN) as the indicator of regularity of orbits. Concretely, the SN is defined to be the number of peaks over a given threshold in the power spectrum of a specific variable that can represent some basic properties of the orbital motion. The method of SN, which could reflect the long term stability within a relatively short integration time, was introduced by \cite{micht95} and has been successfully applied \citep[e.g.,][]{micht02,zhou09,zhou11}. 

In this study, we calculate the power spectrum of the resonant angle of each test particle in the Trojan region, $\sigma=\lambda-\lambda_d$ where $\lambda$ and $\lambda_d$ are the mean longitudes of the test particle and the planet HD141399 d. After some test runs, we set our integration time to be $4.6\times 10^6$\,yr to cover at least two complete periods of all the secular variation terms in the system. 
 
With the same initial conditions as the one we set for obtaining the maximum eccentricity ($e_{\max}$) map, we perform two runs of calculations and two dynamical maps are obtained, as shown in Figure~\ref{fig:dynmap}. The SN in logarithm is indicated by colours. The red and blue imply the most irregular (chaotic) and most regular (stable) motions, respectively, while the intermediates are shown in the colour bar. The empty area (white in colour) are those orbits that cannot survive our integration time. In the left panel, all planets in this system are nearly coplanar with small inclination of $0.1^\circ$, and in the right panel we show the dynamical map for the non-coplanar system, in which the planets in the HD141399 system from inside out are assumed to have the same inclinations and ascending nodes as Jovian planets in our Solar System, i.e. Jupiter, Saturn, Uranus and Neptune, respectively. We will call the latter one the \textit{Jovian-analogue} system hereinafter. 

\begin{figure}                                                                                                                               
\begin{center}                                                                                                                               
\includegraphics[angle=270,width=8.4cm]{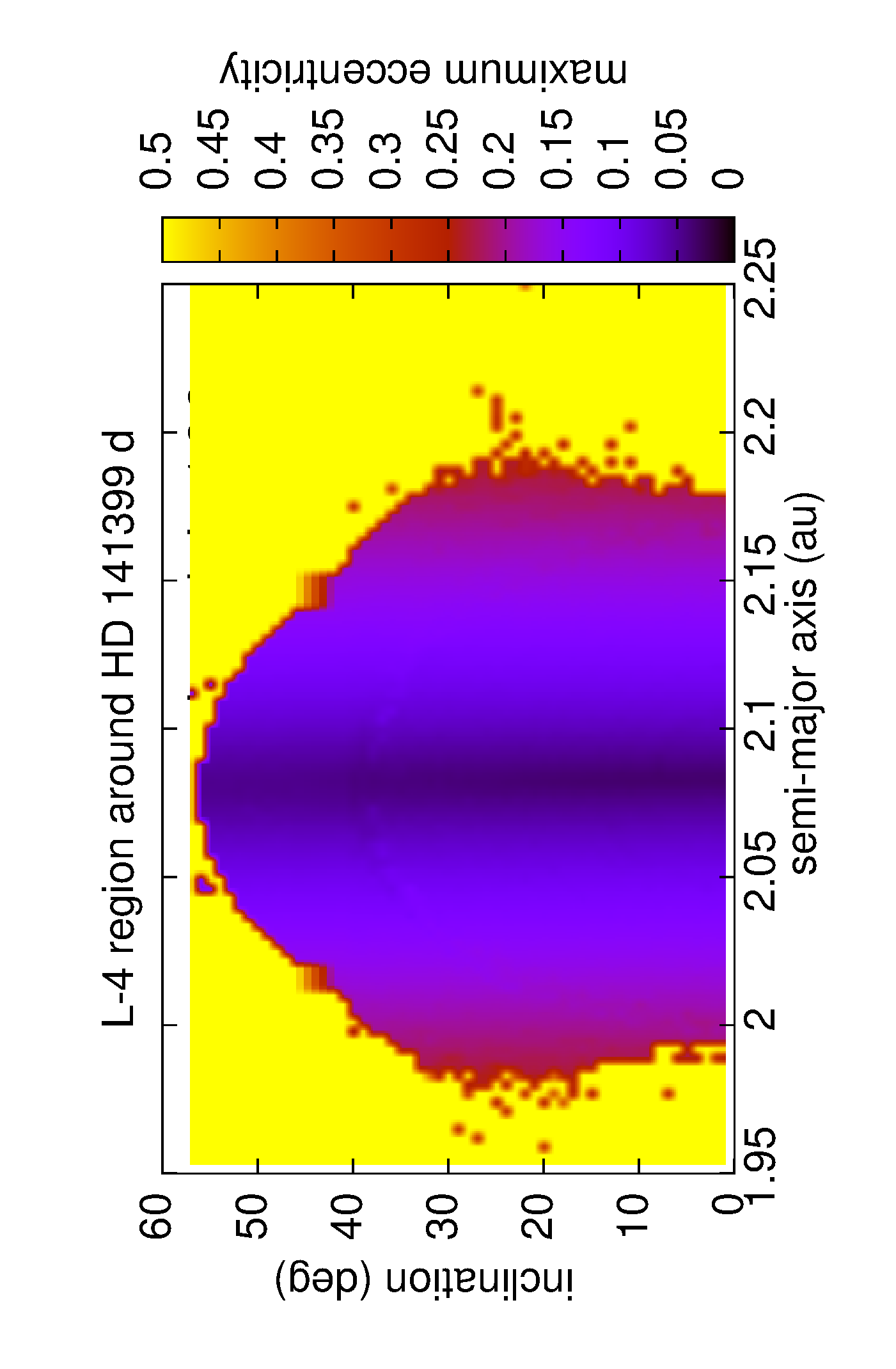}                                                                         
\caption{The results of the stability of orbits in the regions around $L_4$ of HD~141399~d. The initial semi-major axes for fictitious Trojans (x-axes) versus the initial inclination (y-axes) and the maximum eccentricity (z-axes - in color) as stability measure of 6000 orbits. The yellow region outside indicates escape, the black horizontal line is the most stable region.}\label{fig:emax}                                                                                                                                
\end{center}                                                                                                                                 
\end{figure}

The SN dynamical map reveals richer details than the maximum eccentricity map. As shown in Figure~\ref{fig:emax}, the $\mathrm{e_{\max}}$ does not change with the inclination but has a monotonic increase from the resonance centre ($a=2.09$\,au) outwards to both boundaries along the semi-major axis. Contrarily, in Figure~\ref{fig:dynmap} evident arc-structures can be seen, indicating that the orbital stability depends sensitively both on a and i. Introducing chaos into the Trojan region, these distinct structures shrink and shred the stable region. Particularly, the left and right branches of the broad orange arcs corresponding to the most irregular motion are connected to each other at $(a_0,i_0)\sim (2.083\textrm{ au },38^\circ)$, making the stable region at high inclination disconnected from the one at low inclination. 

\begin{figure*}
	\centering
	\includegraphics[width=0.45\textwidth]{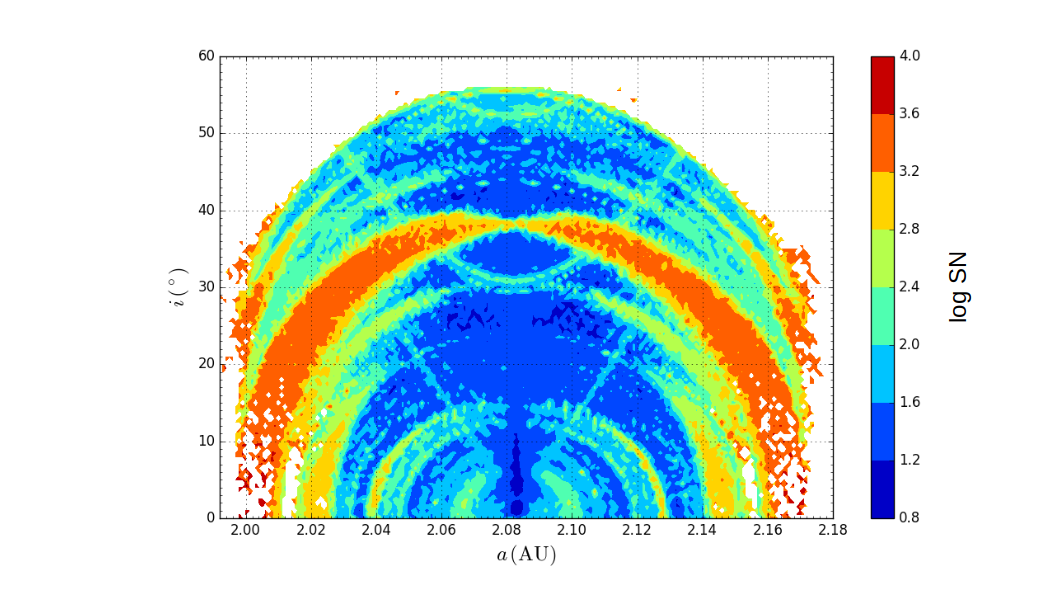}
	\includegraphics[width=0.45\textwidth]{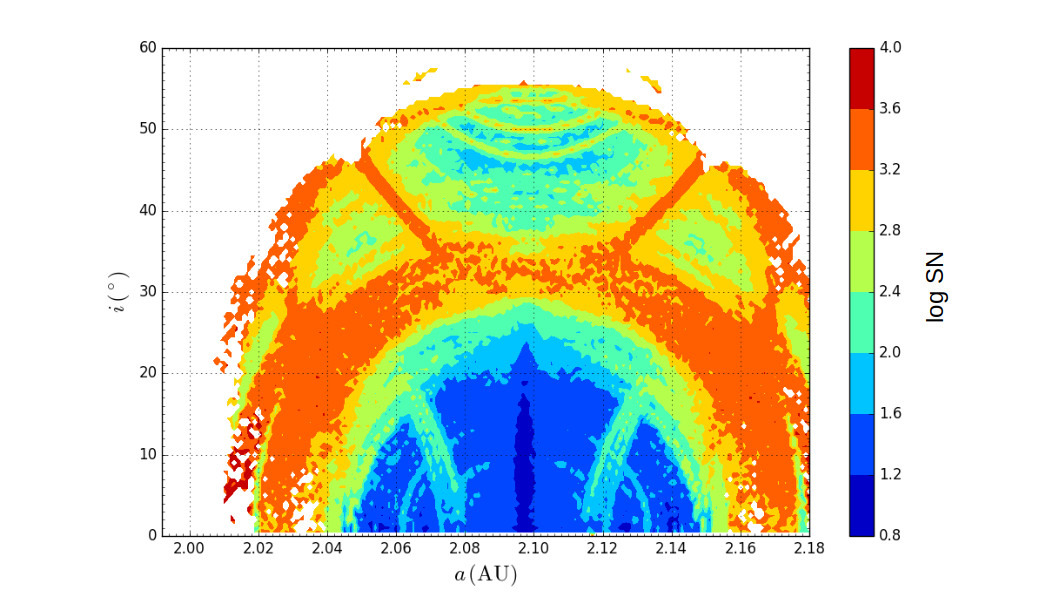}
	\caption{The dynamical map of Trojan region around the $L_4$ point of HD~141399~d. Note the different scales on the x-axes.}
	\label{fig:dynmap}
\end{figure*}

To our experience, the arc structures in the left panel of Figure~\ref{fig:dynmap} arise due to some secondary resonances, where the frequency of $\sigma$ is commensurable to some secular frequencies of the system. In addition, two V-shaped thin stripes can be clearly seen, which most probably is related to the nodal secular resonances. Since the system is nearly coplanar (planets' inclinations are all $0.1^\circ$), these structures may be enhanced if the planets' inclinations are higher. 

In the right panel, we show the dynamical map in the Jovian-analogue system, in which the inclinations of the planets are taken as the same as the Jovian planets (higher than $0.1^\circ$). Apparently, the V-shaped stripes are enhanced as expected. Compared to the left panel, we find that the irregularity has developed considerably. Although the Jovian inclination is still quite small in fact, the stable region in high inclination $\mathrm{i}_0>25^\circ$ almost disappears now. But in the low-inclination region, the stability increases. As a result of the disappearance of those cyan-yellow arc structures corresponding to irregular motion, the blue stable region extends both to lower inclination and to a wider range in semi-major axis. It is also worth to note that the resonance centre in this Jovian-analogue system shifts a little from the near-coplanar system, because here all the initial conditions are osculating orbital elements. 

\subsection{Resonance Map}\label{sec:resonance}

Generally, a \textit{resonance} happens when two (or multiple) frequencies in a dynamical system match each other or they are commensurable by some simple integers. So, it is possible to find the resonances if the frequencies of the system are known. Following the similar method and procedure as in \cite{zhou09, zhou18}, we numerically determine the frequencies of four planets in this system, which are listed in Table~\ref{tab:profre}. 

\begin{table*}
	\centering
	\caption{The proper frequencies of four planets.The $f$, $g$ and $s$ represent the frequencies of mean motion, periastron precession and nodal precession, respectively. The values are all given in a unit of $2\pi/$yr. The sign $\pm$ indicates the precession direction, `$+$' for prograde and `$-$' for retrograde. The nodal precession of Planet d is so slow that we cannot detect it. }
	\begin{tabular}{c|llll}
		\hline
		{ } & ~~~~Planet b & ~~~~Planet c & ~~~~Planet d & ~~~~Planet e \\
		\hline
		$f$ & $+3.866383$ &  $+1.811227$ &  $+3.461740 \times 10^{-1}$ &  $+9.438578\times 10^{-2}$ \\
		$g$ & $+3.751551\times 10^{-3}$ & $+3.406273\times 10^{-4}$ & $+6.832154\times 10^{-5}$ & $+2.066966\times 10^{-5}$ \\
		$s$ & $-2.276786 \times 10^{-3}$ & $-1.240933\times 10^{-4}$ & $0.000000$ & $-3.178531\times 10^{-5}$  \\
		\hline
	\end{tabular}
	\label{tab:profre}
\end{table*}

From Table~\ref{tab:profre}, we notice that the mean motions of Planet c and Planet d are very close to a ratio of 5:1. In fact, in our numerical simulations, we find that this near resonance plays a significant role in the dynamics of Trojans around Planet d, as we will see below. The frequency of this near resonance, defined as the frequency of the angle $(5\lambda_d - \lambda_c)$ and denoted by $f_{5:1}$, is calculated as $f_{5:1}= 8.0 \times 10^{-2} 2 \pi/$yr. 

To find out the resonances that sculpt the features in the dynamical map, we calculate the proper frequencies of the test Trojans on the $(\mathrm{a}_0,\mathrm{i}_0)$-plane as well. For each test Trojan, the frequencies of mean motion $f$, apsidal precession $g$ and nodal precession $s$ are calculated and finally the empirical expressions of these frequencies as functions of the initial conditions $\mathrm{a}_0,\mathrm{i}_0$ are obtained. With all the frequencies in hand, we can determine the resonances on the $(\mathrm{a}_0,\mathrm{i}_0)$-plane by carefully checking the relationships among these frequencies. The most significant resonances are found and their locations are plotted on the $(\mathrm{a}_0,\mathrm{i}_0)$-plane. 

\begin{figure*}
	\centering
	\includegraphics[width=0.50\textwidth]{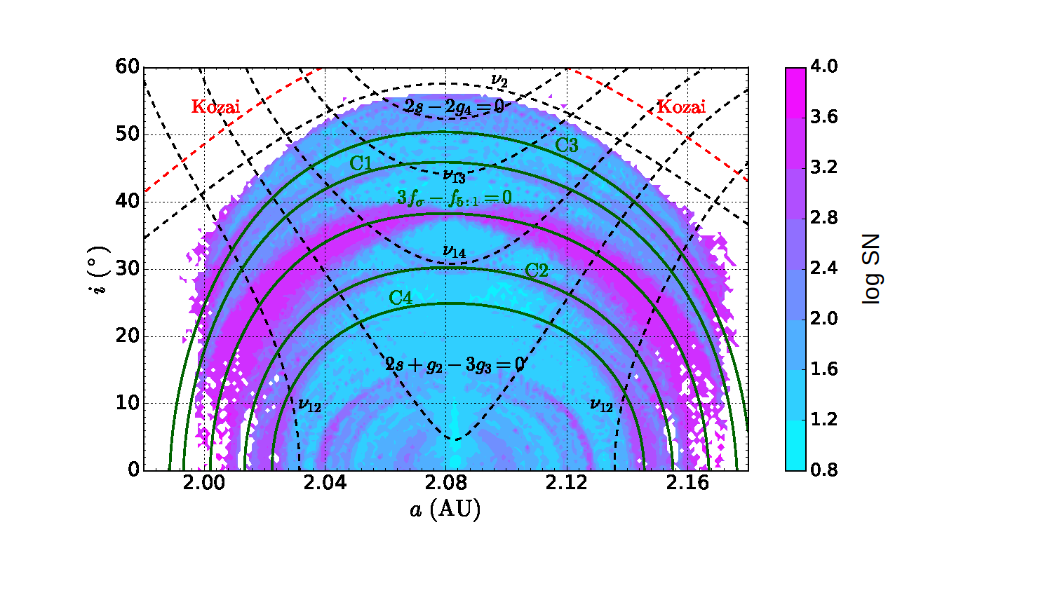}
	\caption{The resonances in the Trojan region around the $L_4$ point of HD141399 d, superposed on the dynamical map, which is the same as the left panel in Figure~\ref{fig:dynmap} but with a different colour code to highlight the resonances. The symbols $\nu_i$ ($\nu_{1i}$), indicating the secular periastron (nodal) resonance between the Trojan and the $\mathrm{i}^{\rm th}$ planet ($\mathrm{i}=1,2,3,4$), are adopted from the notations that are widely used in studies of the Solar system. The lines are labelled by the resonances or the frequencies involved. }
	\label{fig:resmap}
\end{figure*}

The most evident structure on the dynamical map is the family of grand `arcs' crossing the stable region. We find that these arcs are basically related to the secondary resonances between the libration of $\sigma$ (critical angle of the 1:1 MMR) and the near resonance angle $(5\lambda_d - \lambda_c)$. As illustrated in Figure~\ref{fig:resmap}, the line $3f_\sigma- f_{5:1}=0$ just crosses the unstable arc, and some other lines labelled by C1 to C4 that stand for the following equations, emplace the similar structures nearby
\begin{eqnarray}
\label{eqn:c-res}
\begin{aligned}
{\rm C1}:  &  3f_\sigma-f_{5:1}+g_1=0,  \hspace{1cm} & { } 
{\rm C2}:  &  3f_\sigma-f_{5:1}-g_1=0, \\
{\rm C3}:  &  3f_\sigma-f_{5:1}+g_1-s_1=0, { } & { } 
{\rm C4}:  &  3f_\sigma-f_{5:1}-g_1+s_1=0.
\end{aligned} 
\end{eqnarray}

The secular resonance that has the strongest dynamical effects is the $\nu_2$ resonance. As explained above, the $\nu_2$ resonance indicates the situation where the Trojan's periastron processes in the same rate as Planet c's periastron, $\dot{\varpi}=\dot{\varpi}_2$. The $\nu_2$ drives the Trojan's eccentricity to vary in a large range and destablizes its orbit, thus carves the upper boundary of the stable region. The Kozai mechanism, occurring just above the $\nu_2$ curve in Figure~\ref{fig:resmap}, may also contribute to the instability in this neighbourhood. 

Other conspicuous resonances include the nodal secular resonances $\nu_{12}, \nu_{13}$ and $\nu_{14}$. Very clearly, the $\nu_{13}, \nu_{14}$ curves just sit exactly inside the V-shaped stripes. The effects of the lowest order secular resonances, particularly the $\nu_{12}$ and $\nu_{14}$ are greatly enhanced when the inclinations of the planets increase to the Jovian-analogue system (right panel in Figure~\ref{fig:dynmap}), while those high-order secular  resonances, e.g. the one related to $2s+g_2-3g_3=0$, fade away in a relative sense.


\end{document}